\documentclass[fleqn,usenatbib]{mnras}

\usepackage{newtxtext,newtxmath}
\usepackage{pdflscape}

\usepackage[T1]{fontenc}
\usepackage{ae,aecompl}

\usepackage{graphicx}	
\usepackage{amsmath}	
\usepackage{amssymb}	

\title[Astrometric Microlensing by LAWD~37]{A Predicted Astrometric Microlensing Event by a Nearby White Dwarf}

\author[P. McGill et al.]{Peter McGill,$^{1}$\thanks{E-mail: pm625@cam.ac.uk, lsmith,nwe,vasily@ast.cam.ac.uk}
Leigh C. Smith,$^{1,2}$
N. Wyn Evans,$^{1}$
Vasily Belokurov,$^{1,3}$
\newauthor R. L. Smart$^{2,4}$
\\
$^{1}$Institute of Astronomy, University of Cambridge, Madingley Rd, Cambridge CB3 0HA, UK\\
$^{2}$School of Physics, Astronomy and Mathematics, University of Hertfordshire, College Lane, Hatfield AL10 9AB, UK\\
$^{3}$Center for Computational Astrophysics, Flatiron Institute, 162 5th Avenue, New York, NY 10010, USA\\
$^{4}$Istituto Nazionale di Astrofisica, Osservatorio Astrofisico di Torino, Strada Osservatorio 20, 10025 Pino Torinese, Italy
}

\newcommand{\Dl}{{D_{\rm l}}}
\newcommand{\Ds}{{D_{\rm s}}}
\newcommand{\ThetaE}{{\Theta_{\rm E}}}
\newcommand{\Gaia}{{\it Gaia }}

\date{Accepted XXX. Received YYY; in original form ZZZ}

\pubyear{2015}

\begin{document}
\label{firstpage}
\pagerange{\pageref{firstpage}--\pageref{lastpage}}
\maketitle

\begin{abstract}
We used the Tycho-\Gaia Astrometric Solution catalogue, part of \Gaia Data Release 1, to search for candidate astrometric microlensing events expected to occur within the remaining lifetime of the {\it Gaia} satellite. Our search yielded one promising candidate. We predict that the nearby DQ type white dwarf LAWD~37 (WD~1142-645) will lens a background star and will reach closest approach on November~11th~2019 ($\pm$~4~days) with impact parameter $380\pm10$~mas. This will produce an apparent maximum deviation of the source position of $2.8\pm0.1$~mas. In the most propitious circumstance, \Gaia will be able to determine the mass of LAWD~37 to $\sim3\%$. This mass determination will provide an independent check on atmospheric models of white dwarfs with helium rich atmospheres, as well as tests of white dwarf mass radius relationships and evolutionary theory.
\end{abstract}



\section{Introduction}

Einstein's general theory of relativity  predicts that light passing close to a massive object is deflected \citep{Einstein1916}. This later led Einstein to the idea that massive objects can act as gravitational lenses and multiply image background sources~\citep[see e.g.,][for a review]{Schneider1992}. In microlensing, the multiple images are typically separated by milliarcseconds and usually cannot be fully resolved, although the photometric brightening of the source and the astrometric deviation of the light centroid can be in principle detected. \citet{Paczynski1995} noted that microlensing events can be predicted where high proper motion objects (lenses) approach the location of background sources. High proper motion stars are generally nearby and therefore have well-determined distances, which allows the lens mass to be found with high accuracy. The advent of data from the {\it Gaia} satellite, which is providing parallaxes and proper motions for over a billion stars in the Galaxy, makes it timely to look at Paczynski's suggestion anew~\citep[e.g.,][]{Be02,Ha18}.

This {\it Letter} is structured as follows. In section 2, the theory of mass determination via astrometric microlensing is described. Section 3 outlines the methods we used to search for lenses in the Tycho-\Gaia Astrometric Solution (TGAS) catalogue, part of \Gaia
Data Release 1 (DR1) \citep{GaiaA,GaiaB,GaiaAstrometry}. Section 4 gives details of our best candidate event. Finally, in section 5, we sum up with an assessment of the feasibility of observing this event with \Gaia and the Hubble Space Telescope ({\it HST}) and in section 6 summarize the outlook for, and implications of, a precision measurement of the mass of LAWD 37.
 
\begin{figure}
\includegraphics[width=\columnwidth]{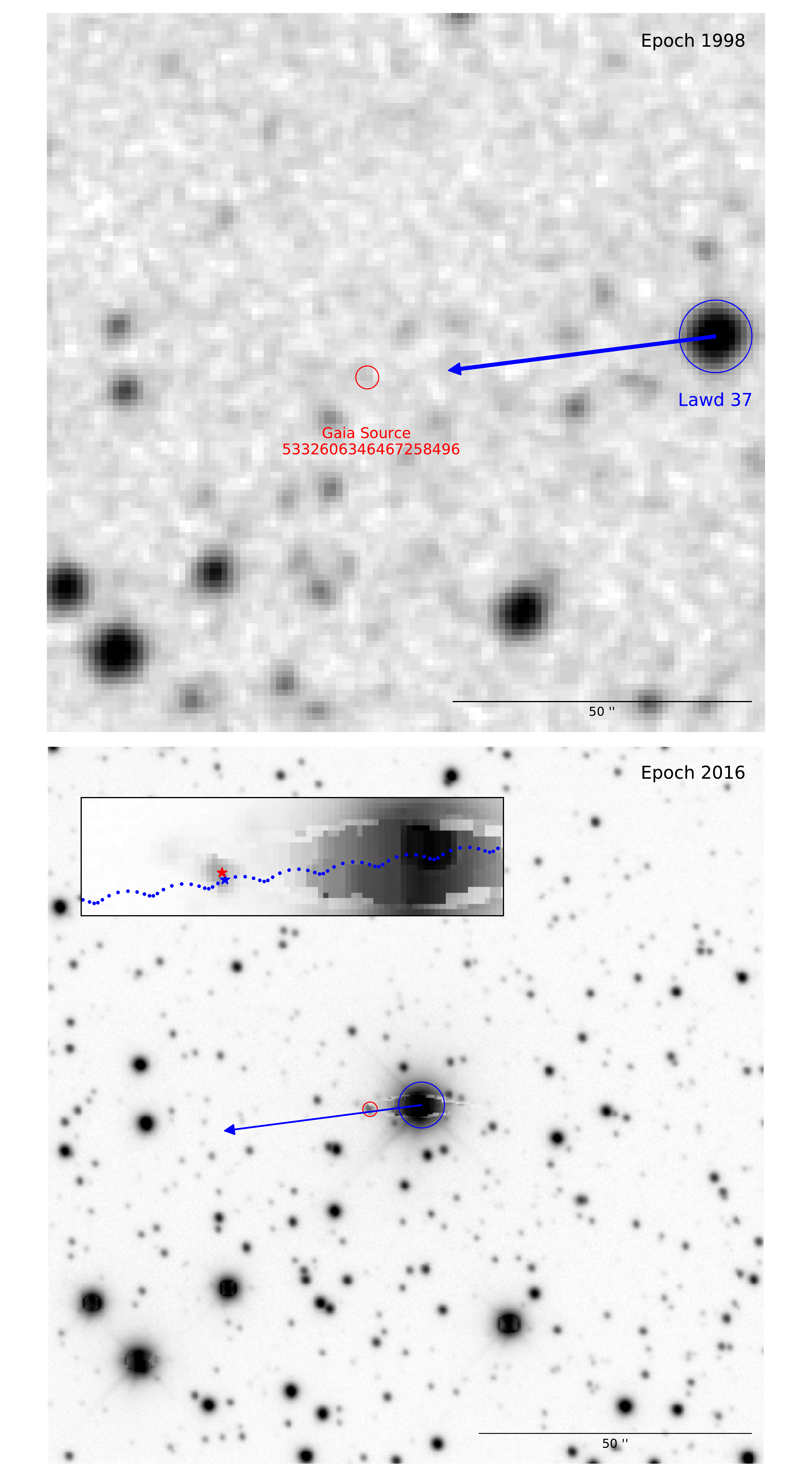}
 \caption{Images of the stellar field around the event. Top: Digitized Sky Survey image at epoch 1998. Bottom: Dark Energy Camera Plane Survey (DECAPS) \citep{Decaps2017} image at epoch 2016. On both images the blue circle indicates the position of the lens LAWD~37, the red circle indicates the position of the source and the blue arrow indicates the proper motion vector of LAWD~37. Bottom inset: Zoom of the DECAPS image. This shows the source and lens position at closest approach marked with red and blue stars respectively. The blue dashed line indicates the lens trajectory which includes parallax motion.}
 \label{fig:stellarfield}
\end{figure}

\section{Mass Determination by Astrometric Microlensing}

Microlensing occurs when a massive point-like foreground object (lens) focuses the light from a background point-like object (source). In the case of perfect alignment between the lens, source and observer, a single Einstein ring with angular radius     
\begin{equation}
		\frac{\ThetaE}{{\rm mas}} \approx 90.2 \text{ } \left(\frac{M}{M_{\sun}}\right)^{1/2}\left(\frac{\text{pc}}{\Dl}\right)^{1/2}, \qquad\qquad \text{if } D_{l} \ll D_{s}
		\label{eq:einstienRadius}
	\end{equation}
is formed. Here, $M$ is the mass of the lens and $\Dl$ and $\Ds$ are the distances to the lens and source respectively. We have assumed that the distance to the source is much greater than the distance to the lens \citep{Paczynski1986}. When a lens encounters a source at some non-zero impact parameter ($\Delta\theta_{\text{min}}$), a bright major image and faint minor image of the source are formed. The major image is located outside the Einstein radius and close to the source, whereas the minor image is located inside the Einstein radius and is close to the lens. In order, the major image, source, lens and minor image always lie along the same line \citep[see][Fig 2.]{Boden1998}. The position of the images relative to the lens are given as~\citep{Paczynski1986}
\begin{equation}
\frac{\theta_{\pm}}{\text{mas}} = \frac{1}{2}\left[\pm\left(u^{2}+4\right)^{1/2}+u\right]\frac{\ThetaE}{\text{mas}},
\end{equation}
where we have defined the dimensionless instantaneous angular separation of the source and lens as $u=\Delta\theta/\ThetaE$. Here, and in the following eqns the positive and negative parts refer to the major and minor images respectively. At closest approach, $u=u_{\text{min}}=\Delta\theta_{\text{min}}/\ThetaE$. The amplifications of the two images are given as \citep{Paczynski1986}
\begin{equation}
A_{\pm} = \frac{u^{2}+2}{2u \left(u^{2}+4\right)^{1/2}}\pm1.
\end{equation}
This amplification causes an apparent brightening of the source (photometric microlensing) and an apparent displacement of the image-source light centroid (astrometric microlensing).  

In the case of a luminous stellar lens and source, in which the lens, source and images cannot be resolved, the apparent centroid shift due to both the major and minor images is suppressed. This is due to light contamination from the luminous lens. The centroid shift is reduced by a factor of $(1+f_{\rm l}/f_{\rm s})$, where $f_{\rm l}$ and $f_{\rm s}$ are the observed fluxes of the lens and source respectively \citep{Dominik2000}. This effect often reduces the astrometric signal by a factor of $\sim100$, making detection difficult~\citep[e.g.,][]{Proft2011}. However, for some events, the impact parameter is large enough that the source and luminous lens can be resolved. In this case, we see an apparent shift of the source centroid, caused by the presence of the major image only. The centroid shift is found by taking the difference between the apparent position of the major image and the true position of the source and is given by \cite{Sahu2017} as
\begin{equation}
\frac{\delta\theta}{\text{mas}} = \theta_{+} - \Delta\theta = \frac{1}{2}\left[\left(u^{2}+4\right)^{1/2}-u\right]\frac{\Theta_{E}}{\text{mas}}.
\label{eq:centroidshift}
\end{equation}
Here, the centroid shift direction is always towards the position of the major image. This is maximal when the lens and source are at closest approach ($u=u_{\text{min}}$). If multi-epoch shifts in the source centroid and lens source separations can be measured for an event, the mass of the lens can be determined using eqns~(\ref{eq:einstienRadius}) and (\ref{eq:centroidshift}), provided that the distance to the lens is known.

\section{Candidate Event Prediction}


To search for events, a high proper motion ($> 150 \text{ mas yr}^{-1}$) sample of $13,206$ lens stars from the TGAS catalog was selected. To narrow our search, the lens sample was cross-matched with the \Gaia DR1 source cataloque. Each lens was paired with all sources within a search radius of 10 times its proper motion. This produced a catalogue of $\sim 4000$ lens-source pairs, which we investigated further by calculating time of closest approach and estimated astrometric deflection. The parallax motion of the lens and the proper motion of the source, where available from the `Hot Stuff for One Year' proper motion catalogue (HSOY) \citep{HSOY2017}, was included.


We define a candidate lensing event as a lens-source pair which has a closest approach within the remaining \Gaia mission time, assumed to be between 2018 and 2022. This left 30 candidate events. Visual inspection of the stellar field around each event removed six suspected erroneous events, which could not be confirmed to be genuine in the images available to us. Of the 24 remaining events, only one had an estimated significant maximum centroid shift in excess of $0.4$ mas. It is this event that we report on here.

\section{The Candidate}


We predict that the known white dwarf LAWD~37 (G magnitude $\sim11$) will encounter a background source (G magnitude $\sim18$) with a closest approach of $\Delta\theta_{\text{min}} = 380\pm10$ mas ($u_{\text{min}}=11.6\pm0.5$) on November 11$^{\text{th}}$ 2019 $\pm$ 4 days ($2019.86\pm0.01$ Julian Years). Fig. ~\ref{fig:stellarfield} shows the stellar field around the event and the trajectory of LAWD~37 as it approaches the source.  The position and proper motion data for both LAWD~37 (the lens) and background source can be found in Table~\ref{tab:astrometricdata}.  Errors in the event parameters were calculated using the uncertainties in source and lens position, proper motion and parallax provided by the TGAS and HSOY catalogues.

\begin{table*}
 \caption{Lens LAWD~37 (first row) and source (second row) data. Proper motions of the lens and source are from the TGAS \citep{GaiaA,GaiaB,GaiaAstrometry} and HSOY \citep{HSOY2017} catalogues respectively. The coordinates ($\alpha,\delta$) are from the \Gaia DR1 source catalogue, on the ICRF and at epoch 2015.0 Julian years. Distance to the lens $D_{l}$ is obtained by inverting the lens parallax of $215.8\pm0.2$ mas from TGAS. G is the {\it Gaia} G band magnitude.}
 \label{tab:LensSourceData}
 \begin{tabular}{ccccccc}
  \hline
  \Gaia DR1 Source Id & $\alpha$ & $\delta$ &$\mu_{\alpha}\cos(\delta)$ & $\mu_{\delta}$ & $D_{l}$ & G  \\ 
                           & [deg $\pm$ mas] & [deg $\pm$ mas] &$[\text{mas/yr}]$ & $[\text{mas/yr}]$ & $[\text{pc}]$ & $[\text{mag}]$ \\
  \hline
  5332606518269523072 & $176.4549073\pm0.2$ &  $-64.84295714\pm0.2$ & $2662.0\pm 0.2$  &  $-345.2\pm 0.2$ & $4.63\pm0.03$ &$11.410\pm0.002$ \\
5332606346467258496 & $176.46360456\pm2$ &$-64.84329779\pm2$ &  $-14\pm3$& $-2\pm3$ & - & $18.465\pm0.005$ \\
  \hline
 \end{tabular}
 \label{tab:astrometricdata}
\end{table*}

At a distance of $\sim4.6$ pc, LAWD~37 (also known as WD 1142-645) is the fourth nearest known white dwarf to the Sun \citep{Sion2009}. It is classified as spectral type DQ indicating the presence of carbon in its atmosphere \citep{Koester1982}. By fitting atmospheric models \citep{Dufour2005} to the photometry of LAWD~37, estimates for its effective temperature ($T_{\text{eff}}=7966\pm219$K) and surface gravity ($\log g = 8.09\pm0.02$) have been obtained \citep{Giammichele2012}. This surface gravity estimate combined with the parallactic distance allows the radius of LAWD~37 to be determined. Assuming LAWD~37 follows the standard evolutionary model for carbon-oxygen (CO) core white dwarfs, \cite{Giammichele2012} estimates that the radius corresponds to a mass of $0.61\pm0.01$ M$_{\sun}$.

\begin{figure}
\includegraphics[width=\columnwidth]{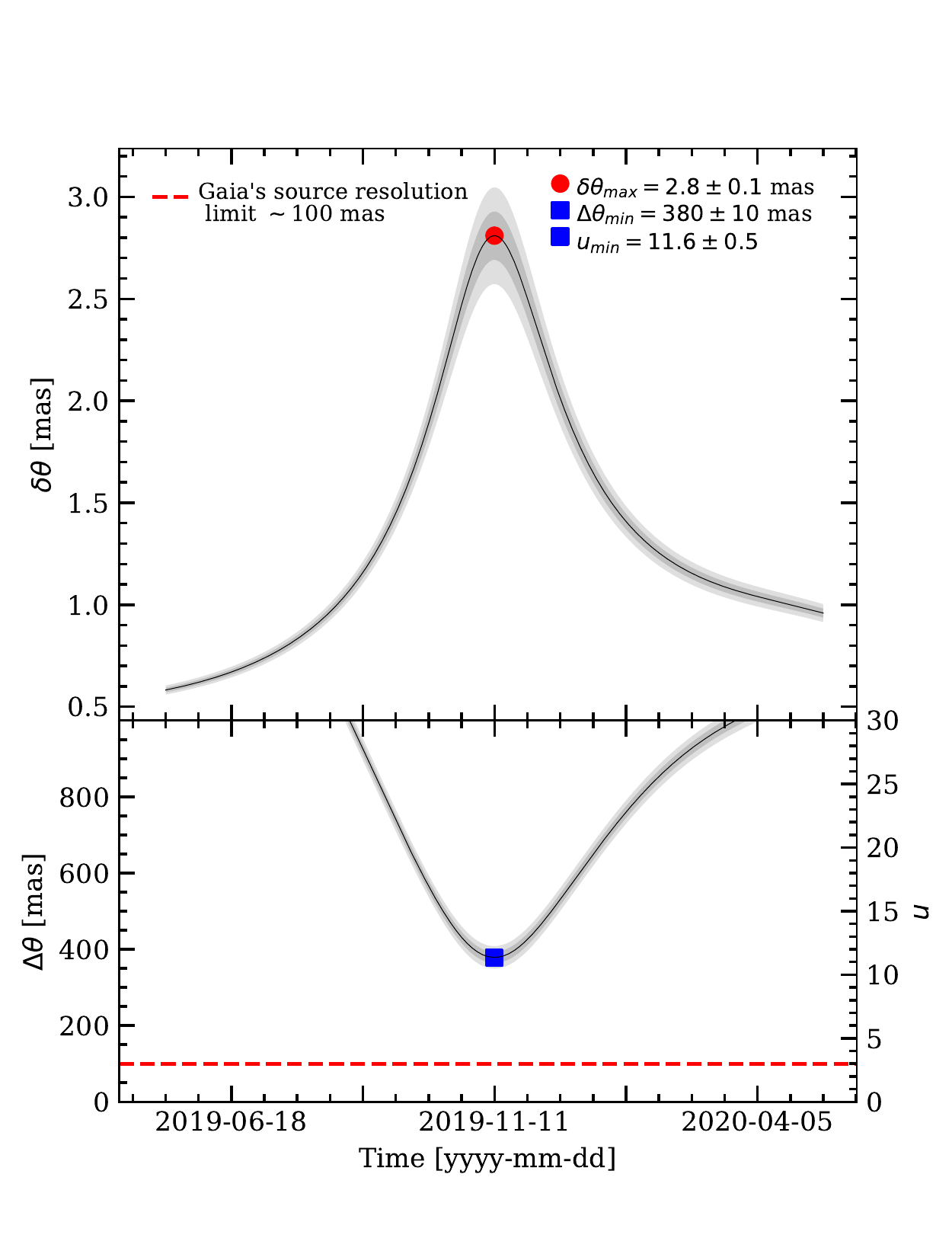}
 \caption{Top Panel: Magnitude of the centroid shift around the time of closest approach ($2019.86\pm0.01$ Julian Years or $2019-11-11 \pm 4$ days) for the event. This is calculated using equation (\ref{eq:centroidshift}). Maximum centroid shift (red circle) is $\delta\theta_{\text{max}}=2.8\pm0.1$ mas. Bottom panel: Lens-source separation around the time of closest approach for the event. The minimum separation (blue square) is $u_{\text{min}}=11.6\pm0.5$, corresponding to a minimum angular separation of $\Delta\theta_{\text{min}} = 380\pm10$ mas. Dark and lighter grey bands indicate 1$\sigma$ and $2\sigma$ errors on both $\delta\theta$ and $\Delta\theta$ respectively. The red dashed line indicates {\it Gaia}'s resolution limit of $\sim100$ mas (see text).}
 \label{fig:astrometricSignal}
\end{figure}

Using eq.~(\ref{eq:einstienRadius}), the mass estimate of \cite{Giammichele2012} and the TGAS parallax, we find the Einstein Radius for LAWD~37 to be $\Theta_{E}=32.8\pm0.3$ mas. We have assumed that the source is sufficiently distant such that $D_{\rm s}\gg D_{\rm l}$. Fig.~\ref{fig:astrometricSignal} shows the estimated astrometric signal and separation of the lens and source during the event. At closest approach, the maximum centroid shift is $\delta\theta_{\text{max}}=2.8\pm0.1$ mas. {\it Gaia}'s resolution limit is a function of the orientation of the objects with respect to the focal plane and the magnitude difference of the objects. However, it is potentially $\sim100$\,mas \footnote{\url{https://www.cosmos.esa.int/web/gaia/science-performance}}, as shown on Fig.~\ref{fig:astrometricSignal}. Due to the event's large impact parameter ($u_{\text{min}}\gg 1$), the photometric signal is estimated to correspond to an apparent maximum brightening of the source of $\sim 10^{-4}$ mag. Therefore the photometric signal is unlikely to be detected by \Gaia, so we consider constraining the mass of LAWD~37 from the astrometric signal only.

\section{Observational Outlook}

With a closest approach of $\Delta\theta_{\text{min}} = 380\pm10$ mas, a predicted astrometric deflection of  $\delta\theta_{\text{max}}=2.8\pm0.1$ mas and a lens-source magnitude difference of $\sim 7$, the viability of successfully observing this event has to be demonstrated.

\subsection{Gaia}

\begin{figure}
\includegraphics[width=\columnwidth]{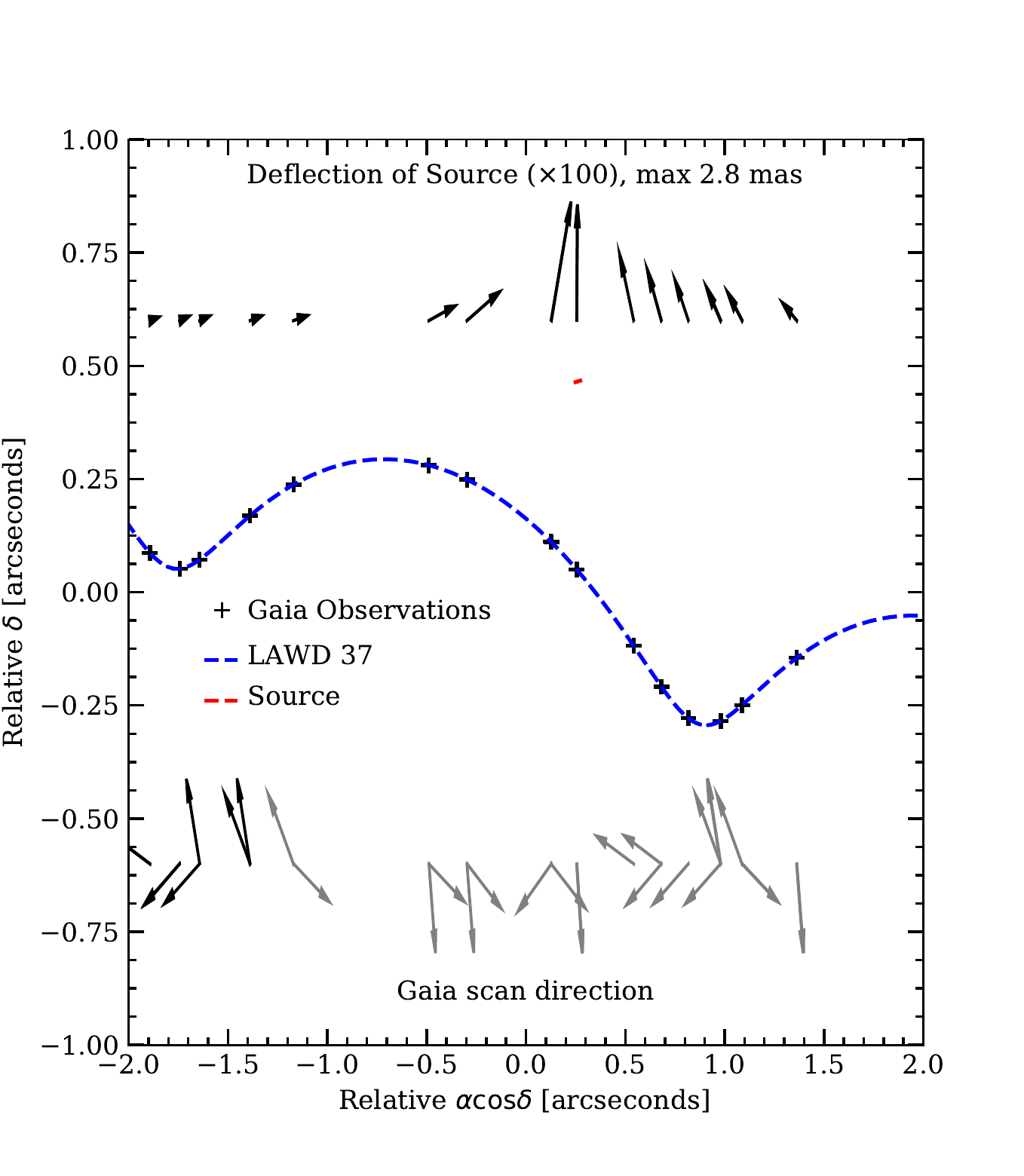}
\caption{Blue and red dashed lines indicates LAWD~37's and the source's trajectory around the time of closest approach. Crosses mark the time of {\it Gaia}'s predicted observations. The vectors at the top indicate the predicted source deflection direction, the largest deflection is $2.8$ mas. The arrows at the bottom indicate {\it Gaia}'s scan directions (gray arrows indicate provisional scan directions after June 2019). When the deflection arrow and {\it Gaia}'s scan direction are aligned, the measurement is along scan and they will be the most precise.}
\label{fig:gost}
\end{figure}

Fig.~\ref{fig:astrometricSignal} shows that the lens and source should be resolvable by \Gaia for the duration of the event. In order to assess the feasibility of observing the event with {\it Gaia}, we use the \Gaia Observation Schedule Tool (GOST)\footnote{\url{https://gaia.esac.esa.int/gost/}} to predict the dates and scan direction of the expected {\it Gaia} observations. Due to the scanning law, observations are unevenly spaced and the scan direction, which is an important predictor of the possible centroiding precision, constantly changes. In Fig.~\ref{fig:gost}, we show the propagated positions of the source and lens when \Gaia is predicted to observe LAWD 37. We have also plotted the scan direction of \Gaia and the estimated direction of deflection.

The final centroiding precision of \Gaia will be determined by a combination of the scan direction and the relative position of the two objects. Particularly for objects fainter than $G$=13, \Gaia provides only binned line-spread-functions with very precise positions along scan directions, but relatively low precision in the across scan direction. For objects as bright as LAWD~37, \Gaia will provide a window 2 x 1'' in the across and along scan respectively \citep{2016A&A...595A...3F}. From
one CCD transit, it is possible to obtain precisions of 0.06 mas for a $G$=12 object \citep{2016A&A...595A...3F}. However, because our primary measurement is the distance between the two objects, the floor will be set by the fainter source. 

When the objects are observed in the same window, the use of gates to stop LAWD~37 saturating will lead to a significantly reduced signal-to-noise of the fainter source and a corresponding loss in precision.  Even when in the same window, the higher precision along scan will remain because the pixels are rectangular and approximately 3 times larger in the across compared to the along scan direction. For the best case scenario, with both objects in the window and aligned along scan, the error on the apparent separation could be lower than 0.2\,mas while in the worse case scenario, with the orientation across scan, the
error could be as high as 1\,mas. This precision will be improved by a factor of 3 as we have 9 independent estimates, one for each column in the focal plane. We simulated a uniform distribution of scan angles and assumed the along and across scan errors above and that the 9 observations provide independent measurements. From this, we find the per epoch median error for the apparent lens source separation is $\sigma_{\rm ls}=0.24$ mas. Current GOST results from around the event maximum indicate that there will be approximately $30$ scans in which the astrometric deflection will be $> 2 \sigma_{\rm ls}$.

Assuming that $\ThetaE = 32.8$ for LAWD~37, we may estimate the precision at which \Gaia could determine its mass. At each \Gaia transit with an expected astrometric deflection $> 2 \sigma_{\rm ls}$, we draw $10^{6}$ samples from a Gaussian centered at the expected deflection and with variance $\sigma^{2}_{\rm ls}$. We have assumed that the error on the true lens source separation is small compared with the error on the apparent lens source separation, so that $\sigma_{\rm ls} \approx{} \sigma_{\text{deflection}}$ since the apparent lens-source separation is the sum of the true lens-source separation and the deflection. Using these samples and inverting eqn (\ref{eq:centroidshift}) for the mass of the lens, we calculate $10^{6}$ simulated measurements for the mass of LAWD~37 at each transit. By taking the mean and variance of the mass measurement distributions for each transit and then calculating the inverse variance weighted average across all transits, we produce a final mass measurement and error. We estimate in the best case that \Gaia should be able to determine the mass of LAWD~37 to $\sim3\%$ precision.



\subsection{Hubble Space Telescope ({\it HST})}

Single measurement accuracies of $\sim$ 0.2\, mas through pointed imaging by the Advanced Camera for Surveys (ACS) and the Wide Field Camera 3 (WFC3) have been achieved with {\it HST} \citep[see eg.][]{Bellini2011}. Although spatial scanning modes with WFC3 have enabled astrometric measurements with a precision 20-40 $\mu$as \cite{Casertano2016}, the magnitude difference in our event precludes this technique (Casertano, private communication). However, for our event the maximum centroid shift is estimated to be $\sim 2.8$ mas, which is well within {\it HST}'s capabilities. Large scale observing campaigns with {\it HST} to constrain masses of single objects via astrometric microlensing are already underway \citep{Kains2017}. Recently, the mass of white dwarf Stein 2015 B was determined with an accuracy $\sim7\%$ via astrometric microlensing \citep{Sahu2017}. This event had a lens-source closest approach $\sim100$ mas. At the point that it was still resolvable by {\it HST} (separation $\sim 500$ mas), this produced a deflection of the background source position of the order of $\sim2$ mas. The Stein 2015 B event is a similar brightness and contrast ratio to the LAWD~37 event and still the deflection was successfully measured by {\it HST}, providing an optimistic outlook for our event.



\section{Discussion and Conclusions}

White dwarfs are comprised mainly of degenerate matter. They are expected to obey a theoretical mass radius relationship (MRR) as they evolve and cool. Observational confirmation of the MRR is problematic, mainly due to the difficulty of determining the mass of white dwarfs. In a small number of cases when a white dwarf is found in an eclipsing or astrometric binary system \citep[see eg.][]{Parsons2016,Liebert2013}, or with a binary main-sequence companion in wide orbit whose radial velocity can be measured independently \citep{2010ApJ...712..585F}, its mass can be calculated. However, for the majority of white dwarfs, the mass has to be determined indirectly using parameters ($T_{\text{eff}},\log g$) derived from atmospheric models. These models are fitted using spectroscopy or broad-band photometry and require assumptions about the interior structure of white dwarfs. Specifically, the thickness of the non-degenerate hydrogen layers usually has to be prescribed, leading to poor constraints on MRRs. For example, \cite{Tremblay2017} mentions that MRRs derived from atmospheric models can vary between 1-15$\%$ depending on whether a thin or thick hydrogen layer is assumed. Additionally, white dwarfs found in eclipsing binaries are post-common envelope, meaning they have interacted with their companion and potentially evolved differently from isolated white dwarfs.


LAWD~37 is a DQ white dwarf, so it has a helium rich atmosphere. This often means thin hydrogen layers are prescribed in the atmospheric models \citep{Tremblay2017}. A mass determination of LAWD~37 by astrometric microlensing is completely independent of atmospheric models. Therefore, in addition to providing an independent check of model assumptions for white dwarfs with helium rich atmospheres, it will provide an important comparison point between theoretical and observed MRRs, and white dwarf evolutionary theory. 

In conclusion, we have predicted that the white dwarf LAWD~37 will lens the light from a background source, causing an apparent deflection of the source position. Maximally, this deflection will be $2.8\pm0.1$ mas on the 11th of November 2019 $\pm$ 4 days. If LAWD~37 and the source are read out in the same window by {\it Gaia}, a mass determination to $\sim 3\%$ precision should be achieved. Recent observations with {\it HST} of a comparable astrometric microlensing event have allowed the successful determination of the mass of white dwarf Stein 2015 B with $\sim7\%$ accuracy. This provides an optimistic outlook for a precision mass determination of LAWD~37 from our event with {\it HST}. 

{\it Gaia}'s second data release (DR2) is set for 25 April 2018. In addition to providing a refined prediction of the event presented in this {\it Letter}, DR2 will likely provide us with the ability to predict a large number of astrometric microlensing events, and hence precise mass measurements of a rich variety of stars.

\section*{Acknowledgments}
PM would like to thank the Science and Technologies Research Council (STFC) for studentship funding. We
would like to thank  \L ukasz Wyrzykowski, Ummi Abbas, Stefano Casertano and Boris G\"ansicke for useful discussions on this work. We would also like to thank the anonymous referee, whose suggestions improved the paper greatly. This work has made use of data from the European Space Agency (ESA) mission {\it Gaia} (\url{https://www.cosmos.esa.int/gaia}), processed by the {\it Gaia} Data Processing and Analysis Consortium (DPAC,
\url{https://www.cosmos.esa.int/web/gaia/dpac/consortium}). Funding for the DPAC has been provided by national institutions, in particular the institutions participating in the {\it Gaia} Multilateral Agreement.





\bibliographystyle{mnras}
\bibliography{refs}


\bsp	
\label{lastpage}

\end{document}